\documentclass[journal,twocolumn]{IEEEtran}
\usepackage{epsfig,makeidx,color}
\usepackage{amsmath,amssymb,bbm}
\usepackage{cite,graphicx,lipsum}
\usepackage{enumerate}
\def\cQ{{\cal Q}}

\def\rT{{\rm T}}

\def\uZ{{\mathbb Z}}

\def\uE{{\mathbb E}}

\DeclareMathOperator*{\argmax}{\arg\!\max}

\newtheorem{mylemma}{\bf Lemma} 

\def\deft{ \buildrel \triangle \over = }

\def\be{ \begin{equation} }
\def\ee{ \end{equation} }
\def\bea{ \begin{eqnarray} }
\def\eea{ \end{eqnarray} }

\def\bq{{\bf q}}

\def\b0{{\bf 0}}

\def\cL{{\cal L}}

\def\cQ{{\cal Q}}

\ifCLASSOPTIONonecolumn
  \newcommand{\figwidth}{0.33\columnwidth}
  
\else
  \newcommand{\figwidth}{0.80\columnwidth}
  
\fi

\begin{document}

\title{Fast Retrial for Low-Latency Connectivity in
MTC with Two Different Types of Devices}

\author{Jinho Choi\\
\thanks{The author is with
the School of Information Technology,
Deakin University, Geelong, VIC 3220, Australia
(e-mail: jinho.choi@deakin.edu.au).
This research was supported
by the Australian Government through the Australian Research
Council's Discovery Projects funding scheme (DP200100391).}}

\maketitle
\begin{abstract}
In this paper, we consider
co-existing two different types of devices
in machine-type communication (MTC), namely type-1 and type-2 devices,
where type-1 devices need short access delay for low-latency
requirements, while type-2 devices are delay-tolerant.
For short access delay, we study the use
of fast retrial in preamble transmissions 
when a group of preambles is divided
into two subsets to support two different types of devices.
Stability conditions are derived using
Foster-Lyapunov criteria in terms of arrival rates, 
the number of preambles, and the number of type-1 devices. 
We also propose an adaptive
algorithm that dynamically decides the
minimum number of preambles 
for type-1 devices under stability conditions.
\end{abstract}


\ifCLASSOPTIONonecolumn
\baselineskip 22pt
\fi

\section{Introduction}

Machine-type communication (MTC)
plays a key role in supporting various 
Internet-of-Things (IoT)
applications within cellular systems
\cite{3GPP_MTC} \cite{3GPP_NBIoT}.
Since it is expected to support a large number of
devices with sporadic traffic in MTC,
random access is considered. 
In particular, slotted ALOHA (which is known to originally stands for
Additive Links On-line Hawaii Area, 
but is now used to mean its random access scheme)
is widely studied for MTC
\cite{Chang15}.

In MTC, for random access, 
an active device (that has data to send)
is to transmit a preamble that
is randomly selected from a pool of preambles, which is shared by
all devices.
Since  the size of the preamble pool is finite,
there exist preamble collision, which happens
if multiple devices choose the same preamble, 
and the performance depends on
the size of the preamble pool. Thus, as in \cite{Choi16}, 
it is considered to adaptively adjust the size of preamble
pool depending on the devices' activity,
while access class barring (ACB) is usually used for access control
with a fixed size of preamble pool in MTC \cite{Jin17}.

There can be multiple types of devices 
with different requirements.
In this paper, we consider two different types of devices, namely
type-1 and type-2 devices (type-1 devices are delay-sensitive,
while type-2 devices are delay-tolerant), that co-exist
in a system and share a pool of preambles.
The main contribution of the paper is two-fold:
\emph{i)}
to support type-1 devices, fast retrial \cite{YJChoi06}
is applied to preamble transmissions for
short access delay (without reserving preambles \cite{Weera19})
and a \emph{sufficient} condition
for the stability (i.e., condition for a finite access delay)
is derived; 
\emph{ii)} an adaptive algorithm 
to stabilize fast retrial for type-1 devices is proposed.
Note that fast retrial is also applied to MTC in \cite{Choi18},
where access control is employed to stabilize
fast retrial. On the other hand, in this paper,
access control (which may not be applicable to delay-sensitive devices
to meet their requirements)
is not used, but dynamic resource allocation is considered.
Furthermore, unlike \cite{Li17_S} \cite{Zhang19} \cite{Thota19}, 
the size of preamble pool for delay-sensitive devices is dynamically adjusted 
without knowing 
the traffic intensity (i.e., arrival rate) of type-1 devices.
Thus, the proposed adaptive algorithm can be used when
the traffic intensity of type-1 devices
is unknown or varying.

\section{System Model}

In this section, we consider a random access system
that consists of one base station (BS) and two different types
of devices that share a group of preambles.

\subsection{Two Different Types of Devices}

Throughout the paper, we consider the case that
two different types of devices, namely types-1 and 2, co-exist.
Type-1 devices need short access delay for low-latency requirements, 
while type-2 devices are delay-tolerant. 
As in \cite{Thota19}, the number of type-1 devices is usually
much smaller than that of type-2 devices,
while type-1 devices need more resources to meet
short access delay requirements.

\begin{figure}[thb]
\begin{center}
\includegraphics[width=\figwidth]{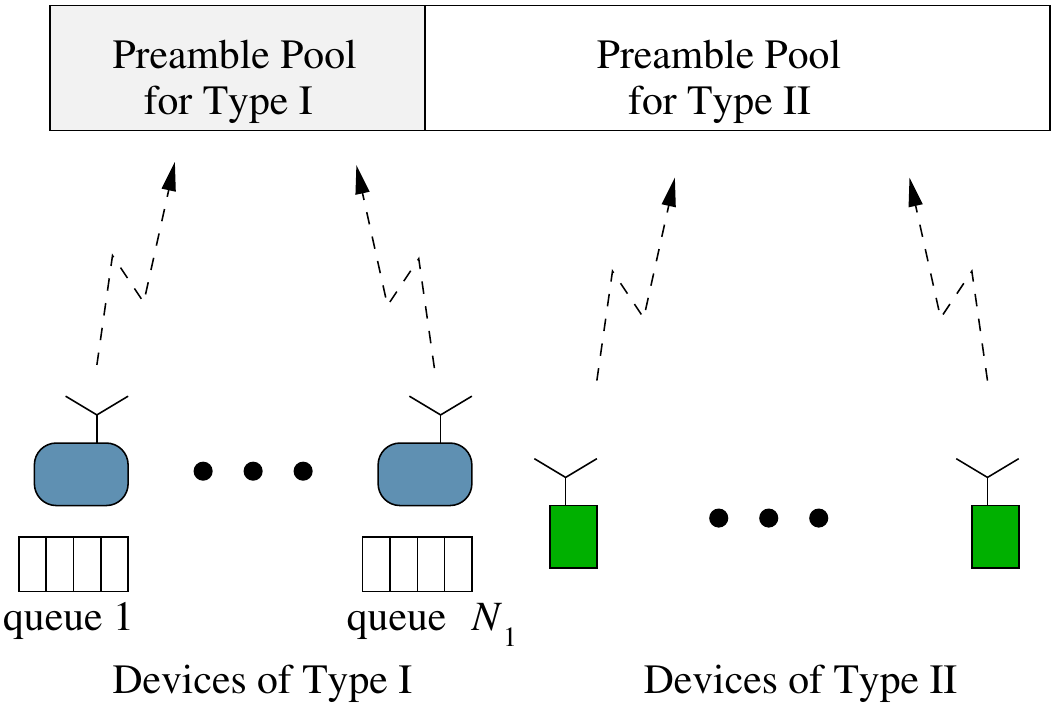}
\end{center}
\caption{An illustration
of two pools of preambles to support two different types of 
devices, where type-1 devices have queues
for re-transmissions.}
        \label{Fig:ra_two}
\end{figure}

A handshaking process as in \cite{3GPP_MTC}
\cite{3GPP_NBIoT} is considered with a group of preambles to support
two different types of devices. 
In particular,
we assume that a group of preambles
is divided into two subsets to support 
two different types of devices 
as illustrated in Fig.~\ref{Fig:ra_two}.
Denote by $\cL_1$ and $\cL_2$
the pools of preambles for type-1 and 2 devices, respectively.
Let $L_i  = |\cL_i|$, $i = 1,2$. Furthermore, let
$L = L_1 + L_2$, which is the total number of preambles.
It is assumed that the BS can adaptively decide
the sizes of the preamble pools.
Thus, $L_1 \in \{1, \ldots, L-1\}$, while $L_2  = L - L_1$.

With a finite size of pool,
since the BS may not be able to detect some transmitted preambles
due to preamble collisions,
it is required for a device 
to re-transmit preambles until a preamble
is successfully
received by the BS according to a re-transmission strategy,
which results in the access delay. Thus, for type-1 devices,
it is important to shorten the access delay for low-latency requirements.

To support type-1 devices with short access delay,
various approaches can be considered.
To this end, in \cite{Li17_S}, 
with two different types of devices, namely
delay-sensitive and delay-tolerant devices
(which might be equivalent
to type-1 and type-2 devices, respectively, in this paper),
$L_1$ can be dynamically adjusted to minimize access delay for 
delay-sensitive devices. 
In this paper, as mentioned
earlier, $L_1$ is also to be dynamically adjusted
for short access delay. In addition, 
we consider fast retrial \cite{YJChoi06}
for type-1 devices with their low-latency requirements.

Note that in \cite{Li17_S}, fast retrial
is implicitly employed for preamble (re-)transmissions.
However, no stability is studied, while we will derive
stability conditions in Section~\ref{S:SQ}.

\subsection{Fast Retrial for Preamble Transmissions}

For short access delay,
an active type-1 device experiencing preamble collision
can immediately re-transmit another randomly selected
preamble in $\cL_1$ in the next time slot
without 
waiting for any back-off time based on fast retrial
\cite{YJChoi06}.

\begin{figure}[thb]
\begin{center}
\includegraphics[width=\figwidth]{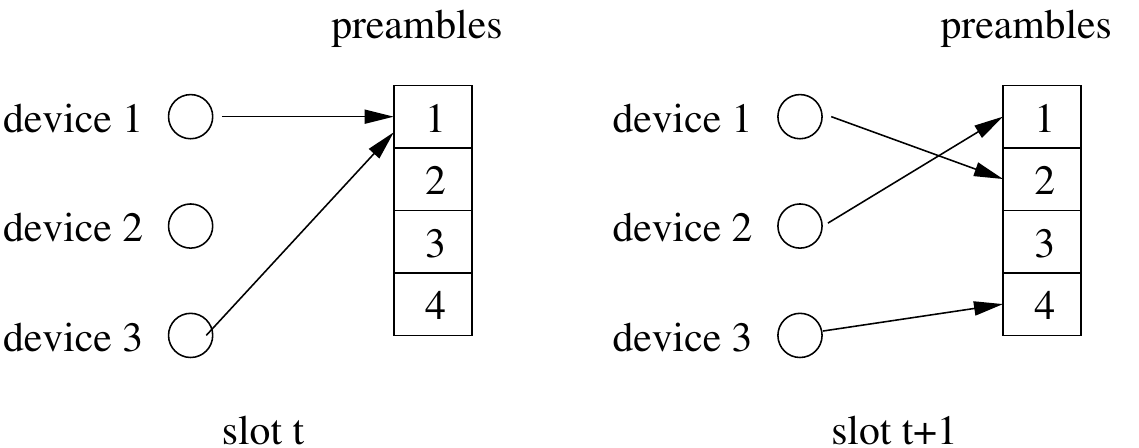}
\end{center}
\caption{An illustration
of fast retrial with 3 (type-1) devices and 4 preambles.}
        \label{Fig:frt}
\end{figure}

In Fig.~\ref{Fig:frt}, we illustrate fast retrial
with $L_1 = 4$ preambles. At slot $t$, suppose that devices 1 and 3
transmit preamble 1, which results in preamble collision.
At the next time slot, i.e., slot $t+1$,
the two devices re-transmit randomly
selected preambles (preamble 2 for device 1 and preamble 4 for
device 3), while a new active device, i.e.,
device 2, transmits preamble 1.
In this case, all the devices can successfully transmit preambles.
This shows that immediate re-transmissions by fast retrial
may not lead to successive preamble collision,
and shorten the access delay
(due to no back-off time).

\section{Stability and Adaptive Algorithm}	\label{S:SQ}

In this section, we will find stability conditions  
for type-1 devices based on
Foster-Lyapunov criteria \cite{Kelly_Yudovina} \cite{HajekBook}.
In addition, an adaptive algorithm to decide $L_1$
is derived.

\subsection{Stability}

Denote by $N_1$ the number of type-1 devices.
In addition,
let $K_1 (t)$ denote the number of active type-1 devices
that send preambles at slot $t$.

At the $n$th type-1 device,
the state of queue is updated as follows:
\be
q_n (t+1) = (q_n (t) + a_n (t) - s_n (t) )^+, \ n = 1, \ldots, N_1,
\ee
where $q_n (t)$, $a_n (t)$, and $s_n(t)$
are the length of queue, the number of new arrivals
(of access request),
and the number of successful preamble transmissions of 
the $n$th type-1 device at slot $t$, respectively.
Here, $(x)^+ = \max\{0,x\}$.
Note that a type-1 device 
becomes active if its queue is not empty, i.e., $q_n (t) > 0$,
with fast retrial.

We assume that $a_n (t)$ is independent and identically distributed 
(iid) with a finite mean as follows:
\be
\lambda_n = \uE[a_n (t)] < \infty,
\ee
where $\lambda_n$ is the average arrival rate
and $\uE[\cdot]$ represents the statistical expectation. 

Since each active device randomly chooses
a preamble from the pool, 
it can be shown that
\be
s_n (t) = 
\left\{
\begin{array}{ll}
1, & \mbox{w.p. $p_n (t)$} \cr
0, & \mbox{w.p. $1 - p_n (t)$} \cr
\end{array}
\right.
\ee
where 
$p_n (t) = \left(1 - \frac{1}{L_1} \right)^{K_1 (t) - 1}$
is the conditional probability of no preamble collision
or successful preamble transmission when there are $K_1 (t)$
active type-1 devices. Furthermore,
it can be shown that
\be
K_1 (t) = | \{n\,|\, q_n (t) > 0\}| \le N_1,
	\label{EQ:KN1}
\ee
because a type-1 device
becomes active when its queue is not empty as mentioned earlier.
As a result, $\bq (t) = [q_1 (t) \ \ldots \ q_{N_1} (t)]^\rT$
is a Markov chain, where $\bq (t) \in \uZ_0^{N_1}$.
Here, $\uZ_0 = \{0, 1, \ldots\}$, i.e., $\uZ_0$
represents the set of non-negative integers.
Since a large $q_n (t)$ means a long queuing delay
for a type-1 device,
it is necessary to avoid. 
To find the conditions for stable queues,
we can consider Foster-Lyapunov criteria
\cite{Kelly_Yudovina} \cite{HajekBook}.

\begin{mylemma}
If
\be
\frac{1}{N_1}
\sum_{n=1}^{N_1} \lambda_n 
< \left(1 - \frac{1}{L_1} \right)^{N_1 - 1},
	\label{EQ:L1}
\ee
$\bq (t)$ is positive recurrent.
\end{mylemma}
\begin{IEEEproof}
Let 
$V(\bq(t)) = \sum_{n=1}^{N_1} q_n (t)$
be a Lyapunov function.
Consider the drift that is defined as
\be
D (\bq) = \uE[V(\bq (t+1)) - V(\bq (t))\,|\, \bq (t) = \bq].
\ee
Let
$\cQ_1 = \{\bq\,|\, q_n \ge 1, \ n=1,\ldots, N_1\}$,
and 
$\bar \cQ_1 = \uZ_0^{N_1} \setminus  \cQ_1$
be the complement of $\cQ_1$.
Thus, for any $\bq$, we have
$\bq \in \cQ_1 \cup \bar \cQ_1$.
Note that $\bar \cQ_1$ is a finite set.
For any $\bq \in \cQ_1$, we have
\begin{align}
p_n (t) = \left(1 - \frac{1}{L_1} \right)^{ ||\bq (t)||_0 - 1} 
= \left(1 - \frac{1}{L_1} \right)^{ N_1  - 1},
\end{align}
where $||\cdot||_p$ denotes the $p$-norm.
Thus, for any $\bq \in \cQ_1$, it can be shown that
\begin{align}
D (\bq) & = \sum_n \uE[a_n (t) - s_n (t)] \cr
& = \sum_n \lambda_n - p_n (t)  = \sum_n \lambda_n - 
\left(1 - \frac{1}{L_1} \right)^{ N_1  - 1}.
\end{align}
Thus, according to \eqref{EQ:L1}, 
we have
\be
D (\bq) \le -\epsilon, \ \bq \in \cQ_1,
	\label{EQ:AL1}
\ee
where $\epsilon > 0$.

For $\bq \in \bar \cQ(N_1)$, there is at least one empty queue
(i.e., $q_n (t) = 0$). For the case of empty queue,
we have
$q_n (t+1) - q_n (t) = (a_n (t) - s_n (t))^+$.
Thus, it can be shown that
\begin{align}
& \uE[q_n (t+1) - q_n (t)\,|\, q_n(t)= q_n]  \cr
& = \lambda_n (1- p_n (t)) 
+ \uE[(a_n (t) - 1)^+ ] p_n (t) \cr
& \le \lambda_n (1- p_n (t)) + \uE[a_n (t)] p_n (t) = \lambda_n ,
\end{align}
which results in
\be
D (\bq) \le \sum_n \lambda_n, \ \bq \in \bar \cQ_1.
	\label{EQ:AL2}
\ee
According to \cite[Proposition D.1]{Kelly_Yudovina},
\eqref{EQ:AL1} and \eqref{EQ:AL2}
imply that $\bq(t)$ is a positive recurrent Markov chain.
\end{IEEEproof}

In \eqref{EQ:L1}, the right-hand side (RHS) term is the 
probability of no preamble collision under full loading (i.e.,
all $N_1$ type-1 devices transmit randomly
selected preambles), which is 
the minimum probability of successful preamble transmission
or the minimum departure rate.
Thus, for a stable system,
\eqref{EQ:L1} 
implies that the average arrival rate on the left-hand side (LHS)
has to be lower than or equal to 
the minimum departure rate.

Let $\lambda_{\rm max}$ be the maximum mean arrival
rate for all type-1 devices so that
$\lambda_n \le \lambda_{\rm max}$. Then,
from \eqref{EQ:L1}, it can be shown that
\be
\lambda_{\rm max} < \left(1 - \frac{1}{L_1}\right)^{N_1 - 1}
\le e^{-\frac{N_1-1}{L_1}}.
\ee
Clearly, from this, with $\lambda_{\rm max} \le 1$, it follows that
\be
N_1 \le \bar N_1 \deft 1 + L \ln \frac{1}{\lambda_{\rm max}},
	\label{EQ:NN}
\ee
where $\bar N_1$
is the maximum number of type-1 devices 
with stable queues or a finite access delay with fast retrial.

For a stable system of type-1 devices,
it is important to decide the key parameters
according to \eqref{EQ:NN}.
Clearly, the number of type-1 devices
has to be less or equal to $\bar N_1$.
In addition, $\lambda_{\rm max}$
can be broadcast to all the type-1 devices
so that their arrival rates cannot be greater
than $\lambda_{\rm max}$.

\subsection{Adaptive Algorithm for $L_1$}


Suppose that $\lambda_n \le \lambda_{\rm max}$.
Then, $L_1$ can be smaller than $L$ 
with a stable system of type-1 devices
so that $L_2 = L -  L_1$ preambles can be
assigned to type-2 devices.
Thus, the minimum $L_1$ that satisfies
\eqref{EQ:L1} has to be found.
Unfortunately, since the $\lambda_n$'s may not be known
to the BS and furthermore the arrival rate
of each device can be time-varying,
the BS needs to estimate $\Lambda  = \sum_n \lambda_n$
to find
the minimum $L_1$. To this end,
we can consider an adaptive algorithm
with an estimate of $\Lambda$.

For convenience, let $z = 1 - \frac{1}{L_1} \in [0, 1)$
and consider
the following function:
\be
f(z) = \frac{1}{N_1} \left(\Lambda z - z^{N_1} \right),
\ee
Clearly,
it can be shown that
$f(z)$ is a concave\footnote{It can be easily shown
that the second derivative of $f(z)$ is 
greater than or equal to $0$.} 
function of $z$
and its derivative becomes
\be
\frac{d f(z)}{dz} = \frac{\Lambda}{N_1} - z^{N_1-1}.
\ee
As a result, $f(z)$ has the unique maximum 
and the solution, which is
\be
z^* = \argmax_{0 \le z < 1} f(z),
\ee
can be found by setting its derivative 
to zero. With $z^*$, it can be readily shown that
$L_1^* = \frac{1}{1 - z^*}$ satisfies
the equality in \eqref{EQ:L1}.

Recall that $K_1 (t)$ is the instantaneous
number of active type-1 devices at slot $t$.
Provided that the queues 
are stable, the total mean 
departure rate has to be equal to
the sum of new arrival rate
and backlogged rate 
(which is the number 
of type-1 active devices with collided preambles per slot).
Let $\Lambda_{\rm d}$ and $\Lambda_{\rm b}$ denote
the total means of departure  and
backlogged rates, respectively.
Then, we have
\be
\Lambda_{\rm d} = \Lambda + \Lambda_{\rm b}.
	\label{EQ:LL1}
\ee
It can be shown that
\begin{align}
\Lambda_{\rm b} = \uE\left[ K_1 (t) 
\left(1 -
\left(1 - \frac{1}{L_1} \right)^{K_1 (t) - 1} \right) \right].
\end{align}
With a sufficiently large $N_1$,
we consider the following Poisson approximation
for $K_1 (t)$:
$$
K_1 (t) \sim {\rm Pois} (0, \Lambda_{\rm d} ).
$$
Then, it follows that
\be
\Lambda_{\rm b} = \Lambda_{\rm d} - \Lambda_{\rm d} e^{- 
\frac{\Lambda_{\rm d}}{L_1}  }
	\label{EQ:LL2}
\ee
Substituting \eqref{EQ:LL2} into \eqref{EQ:LL1},
we have
$\Lambda = \Lambda_{\rm d} e^{-\frac{\Lambda_{\rm d}}{L_1}}$.
Thus, an estimate of $\Lambda$ is given by
\be
\hat \Lambda = K_1 (t)  e^{-\frac{K_1 (t)}{L_1}},
\ee
which leads to the following stochastic gradient 
ascent algorithm to find $z^*$:
\begin{align}
z (t+1) 
& = z(t) + \mu
\frac{d f(z)}{dz} \biggl|_{\Lambda = \hat \Lambda, z = z (t)} \cr
& = z(t) + 
\frac{\mu}{N_1} \left(K_1 (t)  e^{-\frac{K_1 (t)}{L_1 (t)}}
- N_1 z (t)^{N_1-1} \right),
	\label{EQ:zz}
\end{align}
where $\mu$ is the step-size.
Here, 
\be
L_1 (t) = \lceil \max \{1,
\frac{1}{1 - z(t) } \} \rceil.
	\label{EQ:L1t}
\ee


\section{Simulation  Results}

In this section, we present simulation results
to see the performance of type-1 devices in terms of 
queue length. For simplicity,
we assume that $\lambda_n = \lambda$ for all $n$.

\begin{figure}[thb]
\begin{center}
\includegraphics[width=\figwidth]{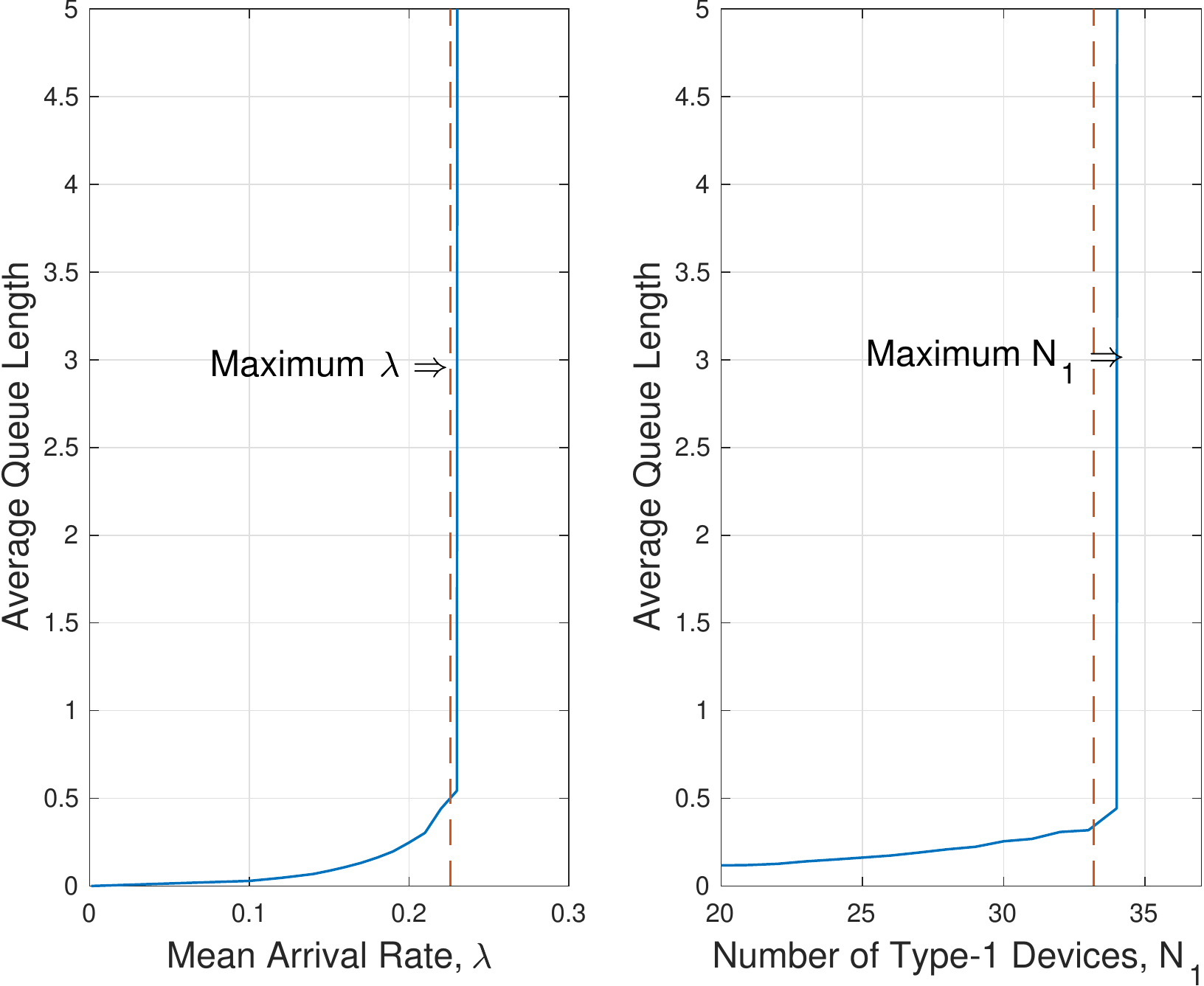} \\
\hskip 0.5cm (a) \hskip 3.5cm (b) 
\end{center}
\caption{The average queue length
when $L_1 = 20$: 
(a) queue length versus $\lambda$ with $N_1 = 30$ and $L_1 = 20$;
(b) queue length versus $N_1$ with $\lambda_n = 0.2$ for all $n$
and $L_1 = 20$.}
        \label{Fig:plt12}
\end{figure}

In Figs.~\ref{Fig:plt12} (a) and (b),
the average queue length, $\uE[q_n (t)]$,
is shown as functions of $\lambda$ 
(with $N_1 = 30$ and $L_1 = 20$) and
$N_1$ (with $\lambda = 0.2$ and $L_1 = 20$), respectively.
Since the access delay increases with
queue length, we can see that
$\lambda$ is to be lower than its maximum,
$\left(1 - \frac{1}{L_1} \right)^{N_1 - 1}$,
when $L_1$ and $N_1$ are fixed (as in Fig.~\ref{Fig:plt12} (a))
or $N_1$ is to be smaller than its maximum,
$1 + L_1 \ln \frac{1}{\lambda}$,
(as in Fig.~\ref{Fig:plt12} (b)) for stable systems.

\begin{figure}[thb]
\begin{center}
\includegraphics[width=\figwidth]{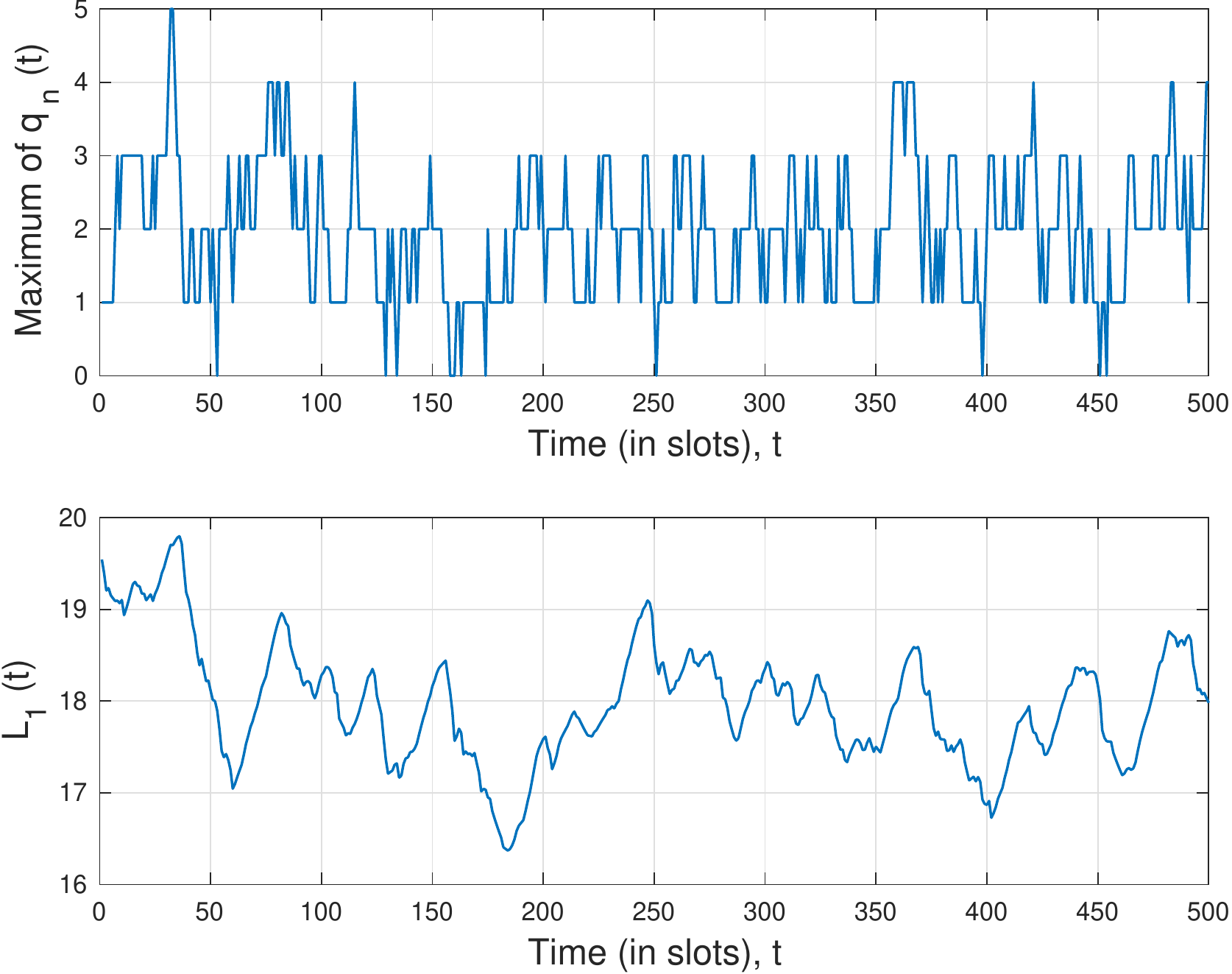}
\end{center}
\caption{Maximum queue length, $\max_n q_n (t)$, and
the number of preambles for type-1 devices
from the adaptive algorithm when $\lambda_n = 0.2$
for all $n$, $N_1 = 30$, and $\mu = \frac{0.01}{N_1}$.}
        \label{Fig:d_all}
\end{figure}

Fig.~\ref{Fig:d_all}
shows the maximum queue length, $\max_n q_n (t)$,
and $L_1 (t)$ in \eqref{EQ:L1t} when
the adaptive algorithm in \eqref{EQ:zz}
is used to dynamically decide $L_1 (t)$
when $\lambda = 0.2$, $N_1 =  30$, and $\mu = \frac{0.01}{N_1}$.
Since the maximum  queue length is finite,
the access delay is also finite. In addition,
if $L = 50$, we can see that more than a half of preambles
can be used for type-2 devices.

\begin{figure}[thb]
\begin{center}
\includegraphics[width=\figwidth]{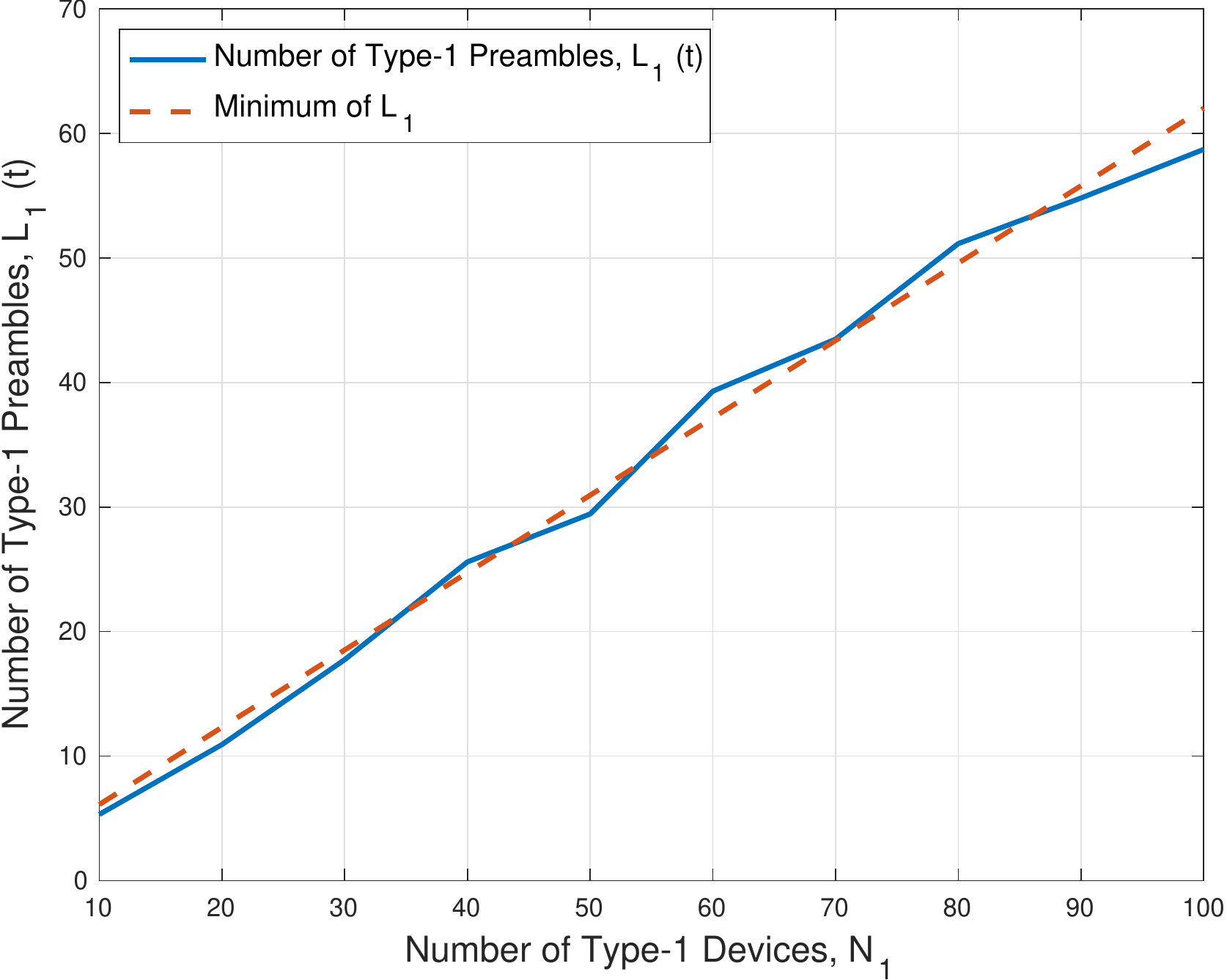}
\end{center}
\caption{The number of preambles for type-1 devices
from the adaptive algorithm for different numbers of type-1 devices
when $\lambda = 0.2$ and $\mu = \frac{0.01}{N_1}$.}
        \label{Fig:plt3}
\end{figure}

Fig.~\ref{Fig:plt3}
shows the average
number of preambles for type-1 devices, $L_1 (t)$,
for different number of type-1 devices, $N_1$,
when $\lambda = 0.2$ and $\mu = \frac{0.01}{N_1}$.
The  minimum $L_1$  is
also shown as a dashed line.
It seems that the adaptive algorithm provides an overestimate of $L_1$.
As a result, a reasonably short queue length
can be achieved (the average queue length is around 2
for a wide range of $N_1$ from simulation results).

\section{Concluding Remarks}	\label{S:Con}

In this paper, we applied fast 
retrial to preamble transmissions for
type-1 devices that require short access delay
when two different types of devices co-exist.
For stable  fast retrial, stability conditions
have been derived using Foster-Lyapunov criteria.
In addition, an adaptive algorithm to decide
the size of preamble pool was derived.

To guarantee a certain access delay,
access control can also be used
together with the adaptation of the size of preamble pool,
which might be a further research topic to
be studied in the future.

\bibliographystyle{ieeetr}
\bibliography{mtc}

\end{document}